\documentclass[a4paper,onecolumn,10pt, noarxiv, accepted=2025-09-09]{quantumarticle}
\pdfoutput=1
\usepackage[utf8]{inputenc}
\usepackage[english]{babel}
\usepackage[T1]{fontenc}
\usepackage{amsmath}
\usepackage{hyperref}
\usepackage{physics}
\usepackage{tikz}
\usepackage{algpseudocode}
\usepackage{algorithm}
\usepackage{lipsum}
\usepackage{amssymb}
\usepackage{bm}

\newtheorem{definition}{Definition}
\newtheorem{proposition}{Proposition}
\begin{document}
\setcounter{MaxMatrixCols}{20}
\title{From Magic State Distillation to Dynamical Systems}

\newcommand{\yun}[1]{\textcolor{red}{#1}}
\author{Yunzhe Zheng}
 \affiliation{Department of Physics, Tsinghua University, Beijing, 100084, China}
  \affiliation{Department of Applied Physics, Yale University, New Haven, Connecticut, 06511, USA}
  \orcid{0000-0002-6419-986X}
  \email{yunzhe.zheng@yale.edu}
\author{Dong E. Liu}
 \affiliation{Department of Physics, Tsinghua University, Beijing, 100084, China}
 \affiliation{Frontier Science Center for Quantum Information, Beijing 100084, China}
 \email{dongeliu@mail.tsinghua.edu.cn}
\maketitle

\begin{abstract}
Magic State Distillation (MSD) has been a research focus for fault-tolerant quantum computing due to the need for non-Clifford resource in gaining quantum advantage. Although many of the MSD protocols so far are based on stabilizer codes with transversal $T$ gates, there exists quite several protocols that don't fall into this class. Here we propose a method to map MSD protocols to iterative dynamical systems under the framework of stabilizer reduction. With the proposed mapping, we are able to analyze the performance of MSD protocols using techniques from dynamical systems theory, easily simulate the distillation process of input states under arbitrary noise and visualize it using flow diagram. We apply our mapping to common MSD protocols for $\ket{T}$ state and find some interesting properties: The $[[15, 1, 3]]$ code may distill states corresponding to $\sqrt{T}$ gate and the $[[5, 1, 3]]$ code can distill the magic state corresponding to the $T$ gate. Besides, we examine the exotic MSD protocols that may distill into other magic states proposed in [Eur. Phys. J. D 70, 55 (2016)] and identify the condition for distillable magic states. We also study new MSD protocols generated by concatenating different codes and numerically demonstrate that concatenation can generate MSD protocols with various magic states. By concatenating efficient codes with exotic codes, we can reduce the overhead of the exotic MSD protocols. We believe our proposed method will be a useful tool for simulating and visualization MSD protocols for canonical MSD protocols on $\ket{T}$ as well as other unexplored MSD protocols for other states.
\end{abstract}

\section{Introduction}

Practical quantum computation must operate at the logical level to achieve robustness to quantum noise \cite{acharyaQuantumErrorCorrection2024,zhouAlgorithmicFaultTolerance2024,guptaEncodingMagicState2024}. To mitigate undesirable error propagation, logical operations are typically implemented transversally such that local errors will remain confined to their respective subsystems during the computation. However, a no-go theorem has shown that the transversal logical operations for any quantum error correcting (QEC) codes cannot achieve universal quantum computation \cite{eastinRestrictionsTransversalEncoded2009, zengTransversalityUniversalityAdditive2007}. As Clifford operations are transversal in many common QEC codes \cite{steaneMultipleParticleInterference1996, fowlerTwodimensionalColorcodeQuantum2011a, fowlerSurfaceCodesPractical2012}, we thus need additional non-Clifford resources for universality. Magic states, or non-Clifford states, are therefore critical resources in universal fault-tolerant quantum computation. 

Although magic states cannot be prepared fault-tolerantly in general, \textit{Magic State Distillation} (MSD) \cite{bravyiUniversalQuantumComputation2005} provides a general framework for obtaining high-fidelity magic states by consuming many copies of moderate-fidelity magic states. Provided that the input states meet certain fidelity requirement, we can distill the faulty magic states to approach ideal fidelity at the cost of a huge amount of raw faulty states. This distillation process relies on Clifford operations, quantum measurements, and classical processing, all often assumed to be noiseless in the MSD setting due to their typical fault tolerance.  Since the inception in 2005, the scope of MSD has been expanded by extensive research. For instance, we can distill into various kinds of magic states \cite{howardSmallCodesMagic2016, Distillation_of_nonstabilizer} apart from the most popular $\ket{T}=(\ket{0}+e^{i\pi/4}\ket{1})/\sqrt{2}$ states that can be exploited to inject the $T$ gate. In addition, recently there has been quite plenty of work that provides insight on lowering the overhead of MSD \cite{bravyiMagicstateDistillationLow2012,hastingsDistillationSublogarithmicOverhead2018, krishnaLowOverheadMagic2019, willsConstantOverheadMagicState2024, nguyenGoodBinaryQuantum2024, golowichAsymptoticallyGoodQuantum2024}, which renders the MSD overhead $\gamma$  for raw magic state cost  approaching to zero \footnote{The asymptotic cost for MSD with at least quadratic error suppression is given by $C \propto (\log \epsilon^{-1})^{\gamma}$, which can be understood as the number of raw states needed to produce a target state with error tolerance $\epsilon$. }. Furthermore, progress in experimental quantum hardware has also triggered interest in the practical application of magic states \cite{guptaEncodingMagicState2024, leeLowoverheadMagicState2024, daguerreCodeSwitchingRevisited2024}. Numerous studies have sought to produce higher-quality magic states tailored to specific scenarios \cite{itogawaEvenMoreEfficient2024, chamberlandFaulttolerantMagicState2019a, gidney2024magicstatecultivationgrowing, goto2016minimizing}. 

Still, most of the previous works concentrate on MSD protocols based on stabilizer codes
with transversal $T$ gates. Due to the natural fault-tolerance of transversal gates, a $[[n,k,d]]$  stabilizer code with transversal $T$ gate will render a $n$-to-$k$ MSD protocol for $\ket{T}$ state with order-$d$ error suppression, i.e. the output error $\epsilon_o = O(\epsilon_i^d)$ scales asymptotically with the input error $\epsilon_i$ in the small error limitation. The raw input states are typically modeled as noisy versions of the ideal $|T\rangle$ state under depolarizing noise, $\rho(\epsilon) = (1-\epsilon)|T\rangle\langle T| + \epsilon Z|T\rangle\langle T|Z$, which relies on the fact that $|T\rangle$ is an eigenstate of the Hadamard-like gate $H_{xy} = \frac{X + Y}{\sqrt{2}}$, allowing input states to be symmetrized via $H_{xy}$-twirling without introducing additional noise \cite{bravyiUniversalQuantumComputation2005}.
However, for magic states higher in the Clifford hierarchy \cite{Distillation_of_nonstabilizer}, the twirling process requires additional non-Clifford resources. As a result, the depolarizing noise model may no longer apply, and input states could exhibit biased noise. Moreover, many stabilizer codes that do not support transversal non-Clifford gates, such as the $[[5, 1, 3]]$ code \cite{bravyiUniversalQuantumComputation2005}, the Steane code \cite{reichardtQuantumUniversalityMagic2005}, and other small codes \cite{howardSmallCodesMagic2016}, can still facilitate magic state distillation. The analysis for these protocols cannot be directly regarded as a QEC process, and we need more delicate techniques to fully understand these MSD protocols.


To fill this gap, we propose a method to map MSD protocols to dynamical systems, which allows to analyze any MSD protocols using theory for dynamical systems. Notably, the idea for the mapping  is not brand new and has been briefly touched in \cite{reichardtQuantumUniversalityState2009,jochym2013robustness, campbell2012magic, rallSignedQuantumWeight2017} for specific protocols and under specific conditions. However, our method can account for input states with biased noise without the twirling assumption, and can be used to analyze any MSD protocols in principle. We can calculate the error suppression rate of MSD protocols by analyzing the Jacobian matrix of the mapped dynamical systems and visualize MSD dynamics using flow diagram. Our method is based on the framework of stabilizer reduction \cite{campbellStructureProtocolsMagic2009}, which is applicable to all MSD protocols characterized by stabilizer codes. We show the dynamical systems should all be rational functions defined in a multi-dimensional space. For single-qubit magic states, the dynamical system is defined within the Bloch sphere with coordinate $(x,y,z)$. We demonstrate our mapping by visualizing common MSD protocols for the $\ket{T}$ states. Surprisingly, we find the $[[15, 1, 3]]$ protocol can distill into magic states corresponding to $\sqrt{T}$ gates, though the trajectory is unstable and the distillation efficiency is linear. The $[[5, 1, 3]]$ protocol proposed to distill the state $\ket{F}\bra{F}=(I+(X+Y+Z)/\sqrt{3})/2$ is found to be able to distill the $\ket{T}$ state as well. We then investigate the exotic MSD protocols numerically exemplified in Ref. \cite{howardSmallCodesMagic2016}, and show that the angle for distillable exotic magic states must be a solution for a single-variable polynomial equation. We also demonstrate convenient evaluation for error suppression rate using Jacobian matrix. Finally, we consider the effect of code concatenation for MSD protocols. Assisted by our mapping method, we numerically show that concatenating two codes with different magic states can generate numerous protocols with new magic states. By concatenating efficient MSD protocols with exotic MSD protocols, we may reduce the overhead for exotic MSD protocols despite of the linear suppression. We believe our method will not only be a useful tool for visualizing MSD protocols, but also an important technique for discovering new interesting MSD protocols that distill into more versatile magic states.

\section{Stabilizer Reduction}
In MSD protocols, we typically begin with a stable supply of raw logical magic states that contain moderate noise. These raw magic states can be prepared through various methods, such as non-fault-tolerant unitary encoding circuits \cite{higgottOptimalLocalUnitary2021a, li2015magic} or measured-based state preparation factories \cite{itogawaEvenMoreEfficient2024, chamberlandFaulttolerantMagicState2019a}. Based on a specific measurement pattern, a success post-selection heralds the production of a higher-fidelity magic state. By using the output states as input for subsequent rounds of distillation and applying the MSD protocols recursively, we can distill magic states to an arbitrary low error rate in the asymptotic limit, though which may require a large number of raw states.

There has been various MSD protocols since its proposal in 2005, and most of the known protocols can be fully characterized by stabilizer codes. For these protocols, we can describe them using a class of protocols called \textit{Stabilizer Reduction} \cite{campbellStructureProtocolsMagic2009}:

\begin{definition}
    {\rm (adopted from Ref.~\cite{campbellStructureProtocolsMagic2009})} A $n$-to-$k$ qubit \textbf{stabilizer reduction} (SR) for stabilizer code $\mathcal{Q}$ performs the following: (i) take an $n$-qubit input state $\rho_{in}$; (ii) measure all stabilizer generators $g_i$ for $i=1,2...,(n-k)$; (iii) postselect on the measurement outcomes of error-detecting stabilizers with all $+1$, and record the measurement outcomes of the other non-error-detecting stabilizers to decide the final gauge correction $\mathcal{ C }_g$; (iii) Decode the post-measurement state onto $k$ output qubits with the logical operators $\bar{X}_i$ and $\bar{Z}_i$ for $i=1,2,...k$ and gauge correction $\mathcal{C}_g$.
\end{definition}
If we post-select on all +1 outcome for all stabilizers, the effect of stabilizer reduction is projecting the input state $\rho_{in}$ onto the codespace of $\mathcal{Q}$, and the success probability $p_s$ is given by the overlap of the input state and codespace:
\begin{equation}
    p_s = \Tr[P_{\mathcal{Q}}\rho_{in}P_{\mathcal{Q}}] = \Tr[P_\mathcal{Q}\rho_{in}],
    \label{p_success}
\end{equation}
where $P_{\mathcal{Q}}$ is the codespace projector for code $\mathcal{Q}$ and we used the fact that $P_\mathcal{Q}^2 = P_\mathcal{Q}$. For MSD protocols that allow gauge correction, the overall success rate should be a multiply of $p_s$
\begin{equation}
    p_{s,all} = 2^{n_{g}}p_s,
\end{equation}
where $n_{g}$ is the number of non-error-detecting stabilizer generators. An example of gauge stabilizers is all $Z$-type generators in the $[[15, 1, 3]]$ code \cite{bravyiUniversalQuantumComputation2005}, as only $Z$ errors can be presented in the input magic states if they are under Pauli noise channel\footnote{This is due to Pauli $X$ and $Y$ errors can be effectively regarded as coherent $Z$ errors for $\ket{T}$. }, and $Z$-type generators cannot detect them.  Without loss of generality, we ignore the gauge correction from now and only consider post-selection on all +1 outcome of stabilizer measurements.

\begin{figure*}[ht]
    \centering
    \includegraphics[width=0.9\linewidth]{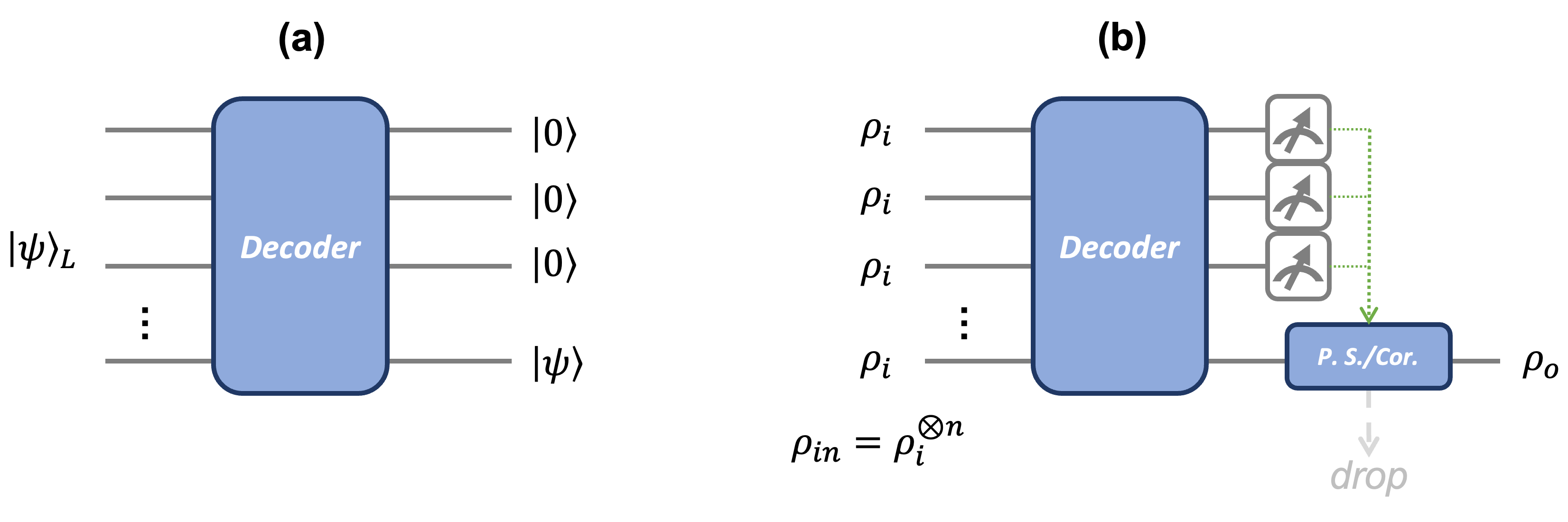}
    \caption{(a) Effect of the decoder operation. The decoder circuit transform the codespace of code $\mathcal{Q}$ to the trivial code with single-qubit stabilizer $Z$ on each ancilla qubit. (b) Scheme for stabilizer reduction protocol characterized by stabilizer code $\mathcal{Q}$ when $\rho_{in}=\rho_i^{\otimes n}$. We prepare the tensor product of raw magic states as the input and apply a decoder circuit for $\mathcal{Q}$. We the measure the $n-k$ ancilla qubits and post-select on the outcome of all +1. For some $\mathcal{Q}$ (e.g. $[[15, 1, 3]]$ code), we only need to post-select on a subset of all stabilizer generators and can correct the state back by using the measurement outcome of the other generators.}
    \label{fig:1}
\end{figure*}

The codespace projector $P_\mathcal{Q}$ is given by the product of $\bar{n}=n-k$ projectors associated with stabilizer generators $g_i$, and can be further expanded as a sum of $2^{\bar n}$ all stabilizer element $s_i$:

\begin{equation}
    P_\mathcal{Q} =\prod^{\bar{n}}_{i=1}P_i =\prod^{\bar{n}}_{i=1}\frac{I+g_i}{2} = \frac{1}{2^{\bar{n}}}\sum^{2^{\bar{n}}}_{i=1}s_i.
    \label{codespace_projector}
\end{equation}
After successful projection, the post-measurement state is given by
\begin{equation}
    \rho_{p} = P_{\mathcal{Q}}\rho_{in}P_{\mathcal{Q}} /p_s \propto P_{\mathcal{Q}}\rho_{in}P_{\mathcal{Q}}.
\end{equation}
At this stage, $\rho_{p}$ is still a $n$-qubit state lying on the logical codespace. For gate injection or next-level distillation, we need to decode the states out with logical operators $\bar{X}_i$ and $\bar{Z}_i$. The output decoded states are given by
\begin{equation}
    \rho^{out}_{i} = \rho(x^o_i,y^o_i,z^o_i) = \frac{1}{2}(I + x^o_iX+y^o_iY+z^o_iZ)
\end{equation}
for $i=1,2..k$, with
\begin{equation}
    \begin{cases}
        x^o_i = \Tr[\rho_{p}\bar{X}_i] = \Tr[P_\mathcal{Q}\rho_{in}\bar{X}_i]/p_s\\
        y^o_i = \Tr[\rho_{p}\bar{Y}_i] = \Tr[P_\mathcal{Q}\rho_{in}\bar{Y}_i]/p_s\\
        z^o_i = \Tr[\rho_{p}\bar{Z}_i] = \Tr[P_\mathcal{Q}\rho_{in}\bar{Z}_i]/p_s,
    \end{cases}
    \label{eq: output_coord}
\end{equation}
where we used the property that all logical Pauli operators should commute with the codespace projector. For convenience, we define
\begin{equation}
    \begin{cases}
    T^x_i \triangleq \Tr[P_\mathcal{Q}\rho_{in}\bar{X}_i]\\
    T^y_i \triangleq \Tr[P_\mathcal{Q}\rho_{in}\bar{Y}_i]\\
    T^z_i \triangleq \Tr[P_\mathcal{Q}\rho_{in}\bar{Z}_i].
    \end{cases}
\end{equation}
Technically, the decoding can be implemented by using a decoder unitary operation $D_\mathcal{Q}$ composed of only Clifford gates and tracing out all ancilla qubits:
\begin{equation}
    \rho^{out} = \Tr_{anc}[D_\mathcal{Q}\rho_p D_\mathcal{Q}^{\dagger}],
\end{equation}
where $\rho^{out}$ is a $k$-qubit state for all $k$ outputs. We could retrieve each output $\rho_i^{out}$  from $\rho^{out}$ by tracing out all outputs except for the $i$th state itself:
\begin{equation}
    \rho_i^{out} = \Tr_{*i}[\rho^{out}].
\end{equation}
The decoding circuit should effectively convert the logical states in codespace $\mathcal{Q}$ back to physical states with ancilla qubits all in $\ket{0}$ (Fig. \ref{fig:1}(a)). For any $k$-qubit logical state $\ket{\bar{\psi}}$, we have
\begin{equation}
    D_\mathcal{Q}\ket{\bar\psi} = \ket{\psi}\otimes \ket{0}^{\otimes \bar{n}}.
\end{equation}
The decomposed implementation of the decoder circuit can be found using Gottesman's algorithm given the stabilizer description of $\mathcal{Q}$ \cite{gottesmanStabilizerCodesQuantum1997}. Besides, as we can propagate the stabilizer measurements after the decoder circuit, we can perform the decoder circuit directly on the input state and measure out every ancilla. The overall circuit implementation of a stabilizer reduction protocol is therefore given in Fig. \ref{fig:1}(b). Notably, there is no need to encode the input state onto the codespace of $\mathcal{Q}$, and only decoding is required.

Almost all known MSD protocols fall under the framework of SR. In \cite{beverland2021cost}, MSD protocols are classified into three types: (a) protocols based on stabilizer codes with transversal non-Clifford gates \cite{bravyiUniversalQuantumComputation2005, bravyiMagicstateDistillationLow2012}; (b) protocols based on stabilizer codespace projection \cite{reichardtQuantumUniversalityMagic2005,bravyiUniversalQuantumComputation2005, howardSmallCodesMagic2016}; (c) protocols based on logical Clifford measurements \cite{jonesMultilevelDistillationMagic2013a, meierMagicstateDistillationFourqubit2012}. It's straightforward to incorporate the first two types to SR protocols. Even though the last type cannot be trivially regarded as SR protocols, it has been shown that these protocols are equivalent to some of the type (a) protocols \cite{haah2018towers}. Therefore, it's sufficient to consider SR protocols in the current scope of MSD. 
\section{Mapping to dynamical System}

We might need multiple times of recursion in MSD to output distilled states with tolerable error rate, and it's a common practice to assume the input state to be a tensor product of identical single-qubit state, i.e. $\rho_{in}=\rho^{\otimes n}$. We now show that we can map the stabilizer reduction protocols into dynamical systems. Without loss of generality, we consider protocols that distill into single-qubit magic states. Any single-qubit quantum state can be represented by
\begin{equation}
    \rho = \rho(x,y,z)=\frac{1}{2}(I+xX+yY+zZ), 
\end{equation}
where $X,Y,Z$ are Pauli matrices and $x^2+y^2+z^2\leq 1$. $x,y,z$ can be understood as the coordinate of the state within the Bloch sphere. Multi-qubit magic states can be represented with coordinates on a higher-dimensional space and will share the same framework as single-qubit states. Although the input state for MSD can be any form in general, we often assume they are the tensor product of the same states due to the recursion nature of the distillation process. In such case, we have:
\begin{proposition}
    When $\rho_{in}=\rho(x,y,z)^{\otimes n}$, $p_s, T_i^x, T_i^y, T_i^z$ are all polynomial of $x, y, z$ with order no larger than $n$.
    \label{prop: poly}
\end{proposition}
\textit{Proof.} Let's show for $p_s$ first. Combine Equation \ref{p_success} and \ref{codespace_projector} and use $\rho_{in}=\rho^{\otimes n}$,  we have
\begin{equation}
    p_s = \frac{1}{2^{\bar{n}}}\sum^{2^{\bar{n}} }_{j=1}Tr[s_j\rho^{\otimes n}].
\end{equation}
Meanwhile, $s_j$ is a $n$-qubit stabilizer operator and can be written as 
\begin{equation}
    s_j = (-1)^{\alpha_j} s^{1}_j \otimes s^{2}_j ... \otimes s^n_j, 
\end{equation}
where $\alpha_j\in\{0, 1\}$ and $s^{l}_j \in \{I, X, Y, Z\}$ for $l=1,...n$. As both the input states and the observable $s_j$ are tensor product of independent single-qubit state (operator), we can make the following simplification and calculate for each single-qubit subsystem
\begin{equation}
\begin{split}
        \Tr[s_j\rho^{\otimes n}] &= (-1)^{\alpha_j} \Tr[ \bigotimes^n_{k=1} s_j^k \rho]\\
        &=(-1)^{\alpha_j}  \prod^n_{k=1}\Tr[s^k_j\rho] \\
        &= (-1)^{\alpha_j} x^{w^X_j}y^{w^Y_j}z^{w^Z_j},
\end{split}
\end{equation}
where $w^X_j, w^Y_j, w^Z_j$ are separately the weight of $X$, $Y$, $Z$ in stabilizer operator $s_j$. Therefore, $p_s$ is a sum of polynomial terms with $x,y,z$:

\begin{equation}
    p_s = \frac{1}{2^{\bar{n} }}\sum_{j=1}^{2^{ \bar n }}(-1)^{\alpha_j}x^{w^X_j}y^{w^Y_j}z^{w^Z_j},
    \label{eq:p_suc_poly}
\end{equation}
and the maximal order of $p_s$ should be no larger than $n$ as $w^X_j + w^Y_j + w^Z_j \leq n$.

Now the same conclusion is easily approachable for $T_i^x, T_i^y, T_i^z$. For example, 
\begin{equation}
    T_i^x = \frac{1}{2^{\bar n }}\sum^{2^{ \bar n}}_{j=1}\Tr[\bar{X}_i s_j\rho^{\otimes n}],
\end{equation}
while $\bar{X}_i s_j$ is still a $n$-qubit Pauli operator. We therefore only need to substitute the $s_j$ to $\bar{X}_i s_j$ in the calculation of $p_s$ and count the Pauli weight for every $\bar{X}_i  s_j$ operator. Therefore, we have
\begin{equation}
    T_i^x = \frac{1}{2^{\bar n}}\sum_{j=1}^{2^{\bar n}}(-1)^{\alpha_j}x^{\Tilde{w}^X_{i, j}}y^{\Tilde{w}^Y_{i,j}}z^{\Tilde{w}^Z_{i,j}},
    \label{eq:T_x_poly}
\end{equation}
where $\Tilde{w}^X_{i,j}, \Tilde{w}^Y_{i,j}, \Tilde{w}^Z_{i,j}$ are separately the weight of $X$, $Y$, $Z$ in stabilizer operator $\bar{X}_i s_j$. The same conclusion also holds for $T_i^y$ and $T_i^z$:
\begin{equation}
     T_i^y = \frac{1}{2^{\bar n}}\sum_{j=1}^{2^{\bar n}}(-1)^{\alpha_j}x^{\bar{w}^X_{i, j}}y^{\bar{w}^Y_{i,j}}z^{\bar{w}^Z_{i,j}}
         \label{eq:T_y_poly}
\end{equation}
\begin{equation}
T_i^z = \frac{1}{2^{\bar n}}\sum_{j=1}^{2^{\bar n}}(-1)^{\alpha_j}x^{\hat{w}^X_{i, j}}y^{\hat{w}^Y_{i,j}}z^{\hat{w}^Z_{i,j}}, 
    \label{eq:T_z_poly}
\end{equation}
with $\bar{w}^X_{i,j}, \bar{w}^Y_{i,j}, \bar{w}^Z_{i,j}$ being the weight of $X$, $Y$, $Z$ in stabilizer operator $\bar{Y}_i s_j$ and $\hat{w}^X_{i,j}, \hat{w}^Y_{i,j}, \hat{w}^Z_{i,j}$ being the weight of $X$, $Y$, $Z$ in stabilizer operator $\bar{Z}_i s_j$.
$\square$

The above proof also provides an explicit way to calculate $p_s, T^x_{i}, T^y_{i}, T^z_i$ for a specific stabilizer code $\mathcal{Q}$. Therefore, for each magic state distillation protocol for single-qubit magic states, we can map it to an iterative dynamical process on a three-dimensional space within the Bloch sphere: 

\begin{proposition}
    For stabilizer reduction protocols that distill into single-qubit magic states, the process can be mapped into a dynamical system within the three-dimensional Bloch sphere. $\rho_{o} = \mathcal{D}_\mathcal{Q}(\rho) \propto Tr_{anc}[D_\mathcal{Q}\bar{P}\rho^{\otimes n}\bar{P}D_\mathcal{Q}]$.
\end{proposition}
For simplicity we now consider $k=1$ case (single output). The output state in Equation \ref{eq: output_coord} is given by
\begin{equation}
    \begin{cases}
        x^o = Tr[\rho_{p}\bar{X}_i] = T_x(x,y,z)/p_s(x,y,z)\\
        y^o = Tr[\rho_{p}\bar{Y}_i] = T_y(x,y,z)/p_s(x,y,z)\\
        z^o = Tr[\rho_{p}\bar{Z}_i] = T_z(x,y,z)/p_s(x,y,z),
    \end{cases}
\end{equation}
and we have a discrete map $(x,y,z)\xrightarrow{\mathcal{D}_\mathcal{Q}} (x^o,y^o,z^o)$ within the Bloch sphere and are able to derive its analytical form. 

Now that we mapped the distillation protocols into dynamical system, we are able to gain several advantages from this mapping: First, We can easily find out all potential target states that a protocol might be able to distill into. This can be done by solving the fixed-point equations and finding every non-trivial fixed points, either analytically or numerically:
\begin{equation}
        \begin{cases}
        x = T_x(x,y,z)/p_s(x,y,z)\\
        y = T_y(x,y,z)/p_s(x,y,z)\\
        z = T_z(x,y,z)/p_s(x,y,z)\\
        x^2+y^2+z^2=1.
    \end{cases} 
\end{equation}
As all target magic states must be fixed points in the dynamical system, we can easily check all potential target states a protocol could distill into. Later we will show this also allows to account for some exotic magic states that distilled by small codes \cite{howardSmallCodesMagic2016}, and all the examples should be a solution of a single-variable polynomial equation. Second, we can obtain the distillation efficiency by analyzing the convergence rate of the fixed points. For example, we can use the Jacobian matrix around a given fixed point $\bm{x}^*=(x^*,y^*,z^*)$ to analysis its first order convergence of all direction. The Jacobian matrix is given by
\begin{equation}
J = 
    \begin{bmatrix}
        \pdv{x}(\frac{T_x}{p_s}) & \pdv{y}(\frac{T_x}{p_s}) & \pdv{z}(\frac{T_x}{p_s}) \\
        \pdv{x}(\frac{T_y}{p_s}) & \pdv{y}(\frac{T_y}{p_s}) & \pdv{z}(\frac{T_y}{p_s}) \\
        \pdv{x}(\frac{T_z}{p_s}) & \pdv{y}(\frac{T_z}{p_s}) & \pdv{z}(\frac{T_z}{p_s})
    \end{bmatrix}_{|\bm{x}^*}.
\end{equation}
 We can calculate the eigenvalue of the Jacobian matrix near a fixed point and judge the property of the fixed point. For discrete system, if $|\lambda_i|<1$ for all the eigenvalue $\lambda_i$ of $J$, then the fixed point is stable and there should exist a nearby convergence region. Furthermore, if they are all zero, we instantly know if the distillation protocols are of at least quadratic error suppression. The eigenvalues also correspond to the prefactor of the convergence speed in a given spatial direction. For higher order analysis, it would be more computationally efficient to determine a spatial direction and then calculate its directional derivative using the analytical description of the system.  Furthermore, the dynamical system also offers the flexibility to consider a more generalized Pauli noise model for the input state. Many previous analyses have focused on depolarized input magic states, which are compatible with twirling operations that symmetrize errors; however, this approach may not be suitable for protocols that distill magic states other than the $\ket{T}$ states. Our framework therefore enables the analysis of input magic states with biased logical noise, which may be suitable for investigating various hardware-specific protocols.

We may further reduce the three-dimensional dynamical system to a two-dimensional system by setting $z=0$ for two reasons: 1. From the perspective of  magic gate injection,  only states formed as $\ket{\theta}=(\ket{0}+e^{i\theta}\ket{1})/\sqrt{2}$ can be injected directly into the logical circuit. Magic states with $z\neq 0$ needs additional post-selection to eliminate the non-zero $z$ component before the injection \cite{bravyiUniversalQuantumComputation2005}. 2.  From the perspective of logical raw state preparation, as the QEC process will ultimately eliminate coherent logical errors \cite{QECdecoheresnoise}
, we might assume that the raw states are the ideal states under logically decoherent Pauli noise. As decoherent noise cannot create the $z$ component of $\ket
{\theta}$, we can make sure that our raw input states are also in the $xy$ plane.  Furthermore, we provide the condition for a stabilizer code such that $z=0$ can be a fixed plane during the MSD process:

\begin{proposition}
    For a stabilizer code $\mathcal{Q}$, $z=0$ is a fixed plane in the mapped dynamical system if $s_j \bar{Z}_i$ explicitly contains Pauli Z operators for every logical operator $\bar{Z}_i (i=1,...,k)$ and every stabilizer element $s_j (j=1,...,2^{\bar{n}})$.
    \label{th: z=0}
\end{proposition}
If each $s_j \bar{Z}_i$ contains at least one Pauli $Z$, then
\begin{equation}
    T^z_i(x,y,z) = z \cdot \tilde{T}^z_i(x,y,z)
\end{equation}
Therefore, $T^z_i(x,y,0) = 0$, which means $z=0$ is a fixed plane for the system. This condition is satisfied by most known protocols that we have examined and we conjecture it to be generally true for any stabilizer codes by properly choosing the logical $Z$ operators without modifying the stabilizer groups.
\begin{figure*}[t]
    \centering
    \includegraphics[width=\linewidth]{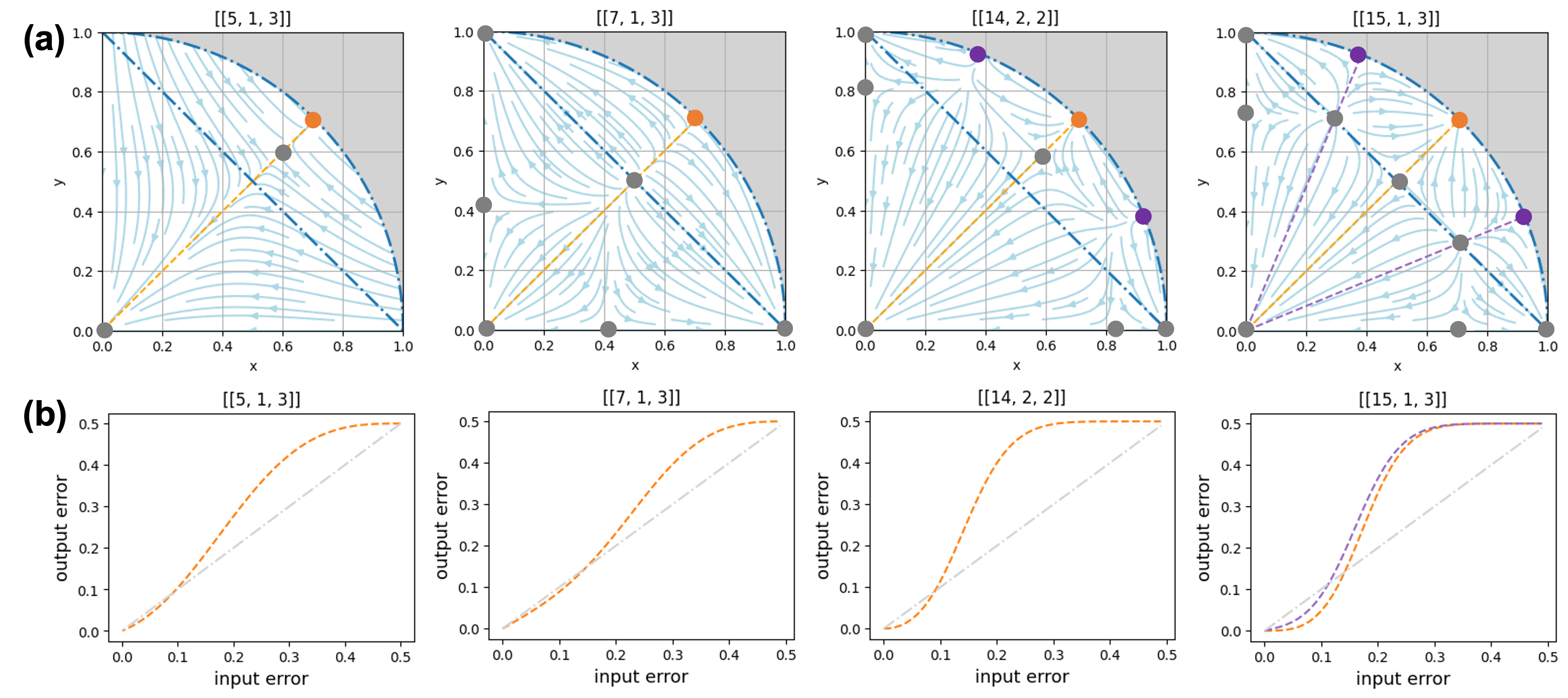}
    \caption{(a) flow diagram in the $z=0$ plane for $[[5, 1, 3]]$, $[[7, 1, 3]]$, $[[14, 2, 2]]$ and $[[15, 1, 3]]$ protocols. All these protocols can distill into the $\ket{T}$ states (yellow dot). The $[[15, 1, 3]]$ can also distill the $\ket{\pi/8}$ and $\ket{3\pi/8}$ states (purple dot). All trivial fixed points are denoted as grey dots.  (b) Input-output error relation for input state under depolarizing noise for each protocol, calculated using the analytical description of the dynamical systems. The plots for the $[[7, 1, 3]]$ and $[[15, 1, 3]]$ protocol distilling the $\ket{T}$ states match the previous results in \cite{bravyiUniversalQuantumComputation2005, reichardtQuantumUniversalityMagic2005}, which is obtained under the specific code setting.}
    \label{fig:ideal_flow}
\end{figure*}

As an initial exploration of our mapping, we simulated the MSD protocols based on the $[[5, 1, 3]]$ code \cite{bravyiUniversalQuantumComputation2005}, $[[7, 1, 3]]$ Steane code \cite{reichardtQuantumUniversalityMagic2005},    the $[[14, 2, 2]]$ tri-orthogonal code \cite{bravyiMagicstateDistillationLow2012} and  $[[15, 1, 3]]$ quantum Reed-Muller code \cite{bravyiUniversalQuantumComputation2005} in Fig. \ref{fig:ideal_flow}. We derive the analytical description of the dynamical system for the given MSD protocols using Eqs. \ref{eq:p_suc_poly}, \ref{eq:T_x_poly}, \ref{eq:T_y_poly}, and \ref{eq:T_z_poly}. Additionally, we analyze the reduced dynamical system by constraining $z=0$, as outlined in Prop. \ref{th: z=0}. The analytical description of the dynamical system for the specific codes can be found in Appendix \ref{app: analytical_descption}. 
 Importantly, although all these MSD protocols can distill into the $\ket{T}$ states, only the $[[14, 2, 2]]$ and $[[15, 1, 3]]$ admit transversal implementation of the $T$ gates. The order of error suppression for both $[[5, 1, 3]]$ and $[[7, 1, 3]]$ is just linear.

In Fig. \ref{fig:ideal_flow}(a), we plot the flow diagram of the dynamical system on the $x$-$y$ cross section of Bloch sphere for $z=0$. This allows us to instantly visualize all fixed points and dynamics of the MSD protocols. We also plotted the output error rate versus input error rate for input state under depolarizing noise in Fig. \ref{fig:ideal_flow}(b), which matches the numerical result from the previous work \cite{bravyiUniversalQuantumComputation2005,reichardtQuantumUniversalityMagic2005}. Actually, we can show the recursive error relation can be obtained from the analytical description of our dynamical systems as presented in Appendix \ref{app: analytical_descption}. Surprisingly, we found out that both the $[[15, 1, 3]]$ codes and $[[14, 2, 2]]$ codes admit $\ket{\pi/8}$ as well as the $\ket{3\pi/8}$ as their fixed points, and $[[15, 1, 3]]$ codes can distill into these magic states by choosing the input states exactly as the depolarized magic states (on the purple dashed line). However, the distillation dynamics is unstable and any perturbation will lead a different output states asymptotically, and the convergence rate is linear. Besides, the Steane codes distilling the $\ket{T}$ states is also unstable and fragile to perturbation. However, this might be circumvent by Hadamard twirling \cite{bravyiUniversalQuantumComputation2005, reichardtQuantumUniversalityMagic2005}, which allows us to engineer the input state to the form of depolarized $\ket{T}$ states on the yellow dashed line. We also find $[[5, 1, 3]]$ code can distill into the $\ket{T}$ state, although it is proposed to distill the magic state $\rho_F = (I+(X+Y+Z)/\sqrt{3})/2$. 

Within the framework of dynamical system, we can also obtain the asymptotic convergence rate as well as the prefactor. For instance, we can calculate the Jacobian matrix for the Steane code at $x=y=1/\sqrt{2}$ based on the analytical form of the system (Appendix \ref{app: analytical_descption}),
\begin{equation}
    J_{*} = 
    \frac{1}{9}
    \begin{bmatrix}
        14 & -7 \\
        -7 & 14
    \end{bmatrix}.
\end{equation}
$J_*$ has two eigenvalues: $7/9$ for eigenvector $\begin{bmatrix}  1 & 1  \end{bmatrix}^T$ and $7/3$ for eigenvector $\begin{bmatrix}  -1 & 1  \end{bmatrix}^T$. This first (radial) eigenvalue matches the prefactor for linear convergence in \cite{reichardtQuantumUniversalityMagic2005}, which shows that the convergence for Steane code is $\epsilon' = \frac{7}{9}\epsilon$ for input state suffering depolarizing noise. The other value is larger than 1, which denotes that the fixed point is not stable along the tangent direction.

Notably, our flow analysis can be done for the whole Bloch sphere without reduction. For comprehensive understanding, We analyzed flow diagrams with various cross sections other than $z=0$ in Appendix. \ref{app: more_flow}.

\section{Analyzing exotic MSD protocols}

In the paper by M. Howard and H. Dawkins \cite{howardSmallCodesMagic2016}, they highlight several small stabilizer codes that could be used to distill exotic magic states other than the canonical $\ket{T}$ states. However, they just provided numerical evidence for these protocols, and the fundamental reason behind these exotic distillable states is still unclear.  Within our mapping framework, we now show that we can analyze these protocols in a convenient way and understand these exotic magic states.
\label{subsec}
The smallest stabilizer code that are able to distill magic states is a $[[3, 1, 1]]$ code with stabilizer generator $\{XZI, ZXX \}$ and $X_L = IZZ, Z_L = IIX$. Although this code cannot even detect a single error, it can be used to distill into the state that is equivalent to $\ket{\theta} = (\ket{0}+e^{i\theta}\ket{1})/\sqrt{2}$ with $\theta = \arctan \sqrt{(\sqrt{5}-1)/2}$, with linear error suppression. However, it's not clear why this code could distill into this particular state. Applying our framework to this code, we can obtain the reduced dynamical system as \footnote{In the original paper they considered the $xz$ plane. To coordinate with their convention, we reduced the dynamical system for $y=0$.} 
\begin{equation}
    \begin{cases}
            x' = \dfrac{xz+z^2}{1+xz+x^2 z}\\
            z' = \dfrac{x+xz+x^2z}{1+xz+x^2 z}.
    \end{cases}
    \label{eq: 3a}
\end{equation}
As we mentioned earlier, any distillable states must be a solution to the fixed point equation for Eq. \ref{eq: 3a}.
We can reduce the equation as
\begin{equation}
x^2 + x^2z+x^3z = xz^2 + z^3.
\end{equation}
Assume $x=\sin\theta$ and $z = \cos\theta$ and define $t=\tan(\theta/2)$, we have $x = \frac{2t}{1+t^2}, z = \frac{1-t^2}{1+t^2}$
\begin{equation}
    t^8 -2t^7 -2t^6 -6t^5 + 8t^4 + 10t^3 + 10t^2 -2t-1 =0,
\end{equation}
which is equivalent to 
\begin{equation}
    (t^4-4t-1)(t^4-2t^3-2t^2-2t+1) = 0.
\end{equation}
Solve it out, we have $t = (1+\sqrt{5} \pm \sqrt{2(1+\sqrt{5})})/2$ or $t= (1\pm \sqrt{2\sqrt{2}-1})/\sqrt{2}$. The solution corresponds to $\theta =  \arctan(\frac{2t}{1-t^2})$. By restricting $\theta>0$, the only solution is $t =(1+\sqrt{5} - \sqrt{2(1+\sqrt{5})})/2 $, which exactly corresponds to $\theta = \arctan\sqrt{\frac{\sqrt{5}-1}{2}}$.

For all magic states, because the MSD protocols should fulfill "good-in, good-out" property, being fixed points for the dynamical systems is a necessary condition to be distillable. As we can always set $x=\sin\theta, z=\cos\theta$ in the dynamical system, combined with Prop. \ref{prop: poly} we instantly have the following conclusion:

\begin{proposition}
    For any exotic magic states distillable with stabilizer codes $\mathcal{Q}$ formed as $\ket{\theta} = (\ket{0}+e^{i\theta}\ket{1})/\sqrt{2}$, $\theta$ must be $\arctan(\frac{2t}{1-t^2})$ where $t$ is a solution for a single-variable polynomial equation associated with the stabilizer codes $f_\mathcal{Q}(t)=0$.
    \label{prop: sol}
\end{proposition}

We also conjecture that there is some structure hidden in the coefficient of the high-order polynomial equation $f_\mathcal{Q}(t)$ that could be related with the weight enumerators and MacWilliams Identities \cite{shorQuantumAnalogMacWilliams1997, rains2002quantum}. With Prop. \ref{prop: sol}, we are able to find out the corresponding $f(t)$ for all the numerical examples given there. As another two examples, we consider the $[[4, 1, 1]]$ code in the Fig. 3(b) and the $[[6, 1, 2]]$ in Fig. 6(c) from \cite{howardSmallCodesMagic2016}. For the four-qubit code, its stabilizer group is generated by $\{XZII, ZXZX, IZXI \}$ with logical operator $Z_L = IIIX, X_L = IZIZ$. This four-qubit code cannot detect a single error as well for its $Z_L$ being single-weight. By doing the same routine for the three-qubit code, we are able to find out that the polynomial equation to be
\begin{equation}
    t^5-t^4+6t^3+2t^2+t-1=0
\end{equation}
This equation doesn't have radical solution, but we are able to numerically solve it to be $t\approx 0.38296$, which corresponds to $\theta \approx 0.73146$. This matches the numerical result provided in the paper for $x=\sin\theta \approx0.66796$ and $z=\cos\theta \approx 0.7442$.

\begin{figure}
    \centering
    \includegraphics[width=0.9\linewidth]{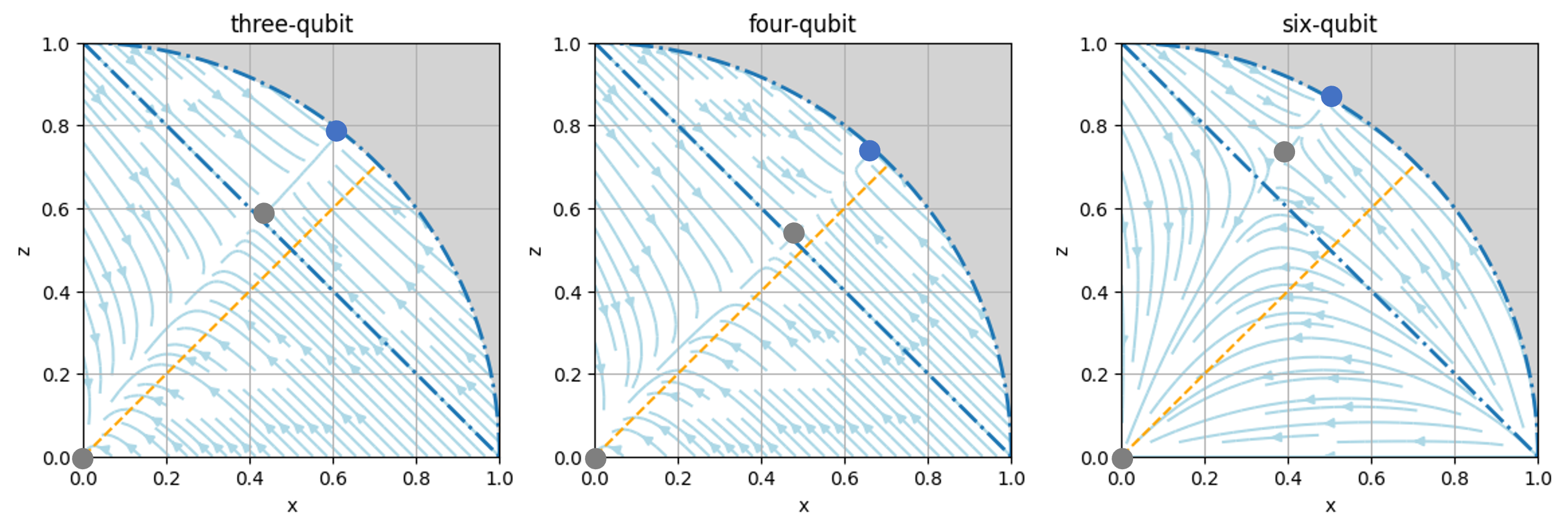}
    \caption{Flow diagram in the $y=0$ plane for the three small codes discussed in Sec \ref{subsec}: $[[3, 1, 1]]$ code, $[[4, 1, 1]]$ code and $[[6, 1, 2]]$ code. The blue dots correspond to the target magic states, and the grey dots are the trivial fixed points. The yellow dashed line corresponds to the $x=z$ for reference.}
    \label{fig:enter-label}
\end{figure}

We can also calculate the asymptotic convergence rate by calculate the Jacobian matrix at the fixed point. The Jacobian matrix at the fixed point $x=\sin\theta_*, z=\cos\theta_*$ can be numerically evaluated: 
\begin{equation}
    J_{|\theta_*} \approx \begin{bmatrix}
        -0.05673 & 0.80093\\
        0.82913 & 0.14814
    \end{bmatrix}
\end{equation}
For this matrix, we have eigenvalue $\lambda_1 \approx 0.867$ and $\lambda_2 \approx -0.775$, with corresponding eigenvector being $[0.867, 1]^T$ and $[-1.114, 1]^T$. As both the absolute value of both eigenvalues are smaller than one, we again confirm that the fixed point is stable. 

For the six-qubit code, it could distill into the $\ket{\pi/6}$ magic states. From the fixed point equation, we can get the expression of the equation $f(t)$:
\begin{equation}
    (t-1)^3(t+1)^5(t^2-4t-1)(t^2-4t+1)=0
\end{equation}
One of the root is $t=2-\sqrt{3}$, which corresponds to $\theta=\pi/6$. The Jacobian matrix is given by
\begin{equation}
    J = \begin{bmatrix}
        38 - 22\sqrt{3} & -48 + 28\sqrt{3} \\
        -62 + 36\sqrt{3} & 118 - 68\sqrt{3},
    \end{bmatrix}
\end{equation}
with eigenvalue $\lambda_1 = 2(5-3\sqrt{3})\approx-0.3923$ with eigenvector $[-\sqrt{3}, 1]^T$ and $\lambda_2 = 2(73-42\sqrt{3})\approx0.5077$ with eigenvector $[\frac{2+4\sqrt{3}}{11}, 1]^T$. The minus eigenvalue corresponds to the direction tangent to the Bloch sphere, while the radial eigenvector has a small angle shift. 

\section{MSD with code concatenation}

As we map MSD protocols to dynamical system, MSD protocols with code concatenation are also mapped to nested dynamical systems. If we have two MSD protocols $\mathcal{Q}_A$ and $\mathcal{Q}_B$ with mapped system $\mathcal{D}_A$ and $\mathcal{D}_B$, then the MSD protocol described by concatenated codes $\mathcal{Q}_A$ (inner) and $\mathcal{Q}_B$ (outer) is mapped to $\mathcal{D}_B \circ \mathcal{D}_A$. The concatenation can be done recursively, which allows to create more MSD protocols from some simple base protocols.
\begin{figure}[H]
    \centering
    \includegraphics[width=
\linewidth]{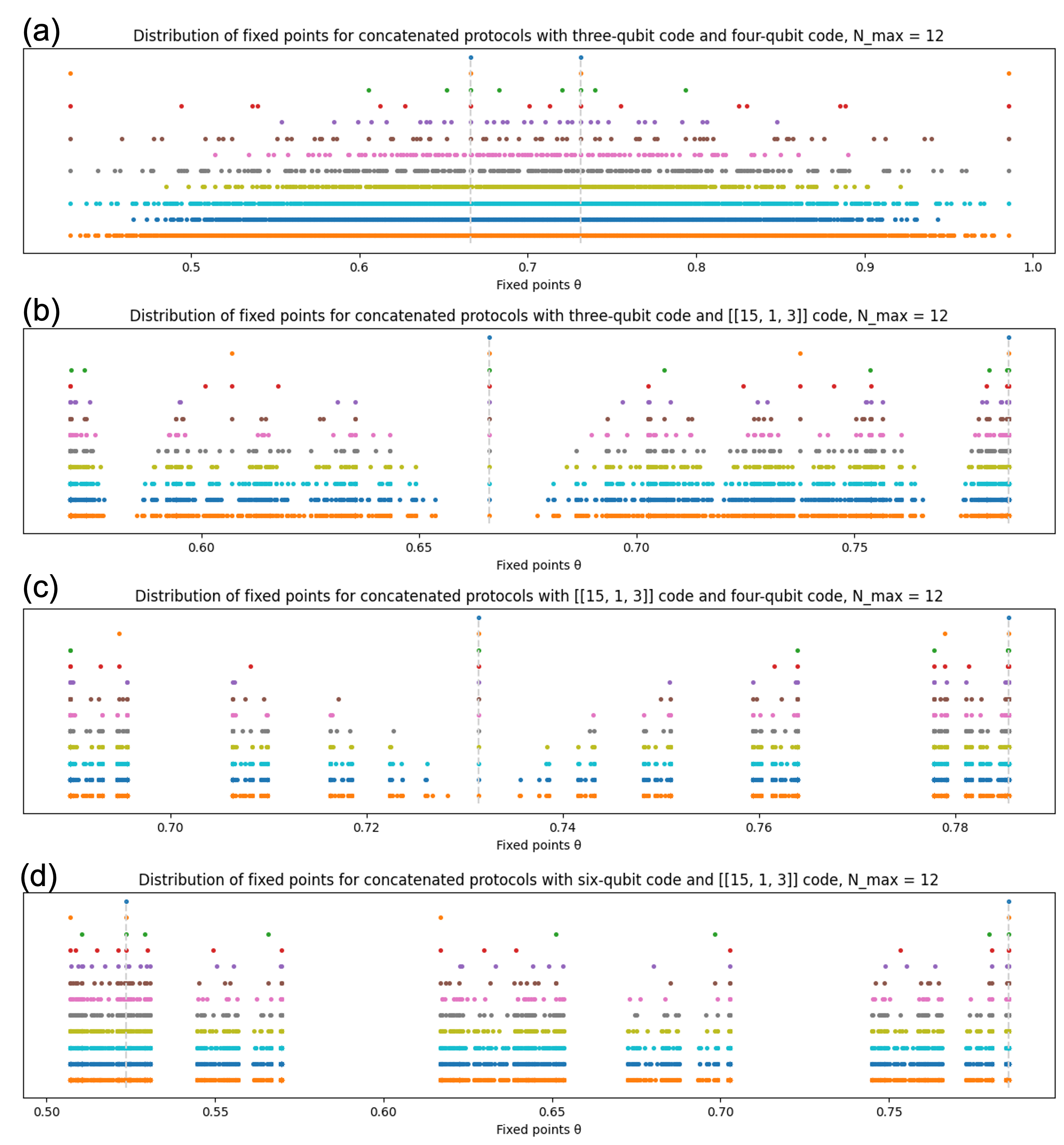}
    \caption{Distribution of fixed point $\theta$ for concatenated MSD schemes, generated from (a) the $[[3, 1, 1]]$ code and the $[[4, 1, 1]]$ code (b) the $[[3, 1, 1]]$ code and the $[[15, 1, 3]]$ code (c) the $[[4, 1, 1]]$ code and the $[[15, 1, 3]]$ code (d) the $[[6, 1, 2]]$ code and the $[[15, 1, 3]]$ code. The maximal level of concatenation increases from 1 to 12 from upper to lower row. The $\theta$ for the two base codes are highlighted with grey dashed lines.}
    \label{fig:distri}
\end{figure}
\subsection{More exotic magic states}

By concatenating codes that distill into different magic states, we can create MSD protocols that distill into new magic states. Take the $[[3, 1, 1]]$ and $[[4, 1, 1]]$ code we mentioned in the last section as the base codes, we can generate various MSD protocols for new magic states with $\theta$ ranging from $0.428$ to $0.985$ if we range the concatenation level from 1 to 12 as shown in Fig. \ref{fig:distri}(a). In the uppermost row, we only consider the $[[3, 1, 1]]$ protocol and $[[4, 1, 1]]$ protocol themselves. By allowing more concatenation level, we can generate a set of MSD protocols with more various target magic states. Notably, the range of generated fixed point angle can be larger than the range of the angles for the two base codes. Besides, for small codes concatenated with $[[15, 1, 3]]$ codes, the distribution of the fixed points exhibit fractal behavior (Fig. \ref{fig:distri}(b-d)), which is typically a result of iterative systems. Notably, fractal behavior for the $[[5, 1, 3]]$ protocol has been studied and close input states might result in total different output states \cite{rallFractalPropertiesMagic2017}. Even though the fractal behavior reported in \cite{rallFractalPropertiesMagic2017} is quite different from our result, it implies the possibility of more fractal behaviors in MSD to be discovered in general. We supplement numerical evidence of the fractal property in Appendix \ref{app: fractal}.

\subsection{Higher distillation efficiency}
\begin{figure}[ht]
    \centering
    \includegraphics[width=0.9\linewidth]{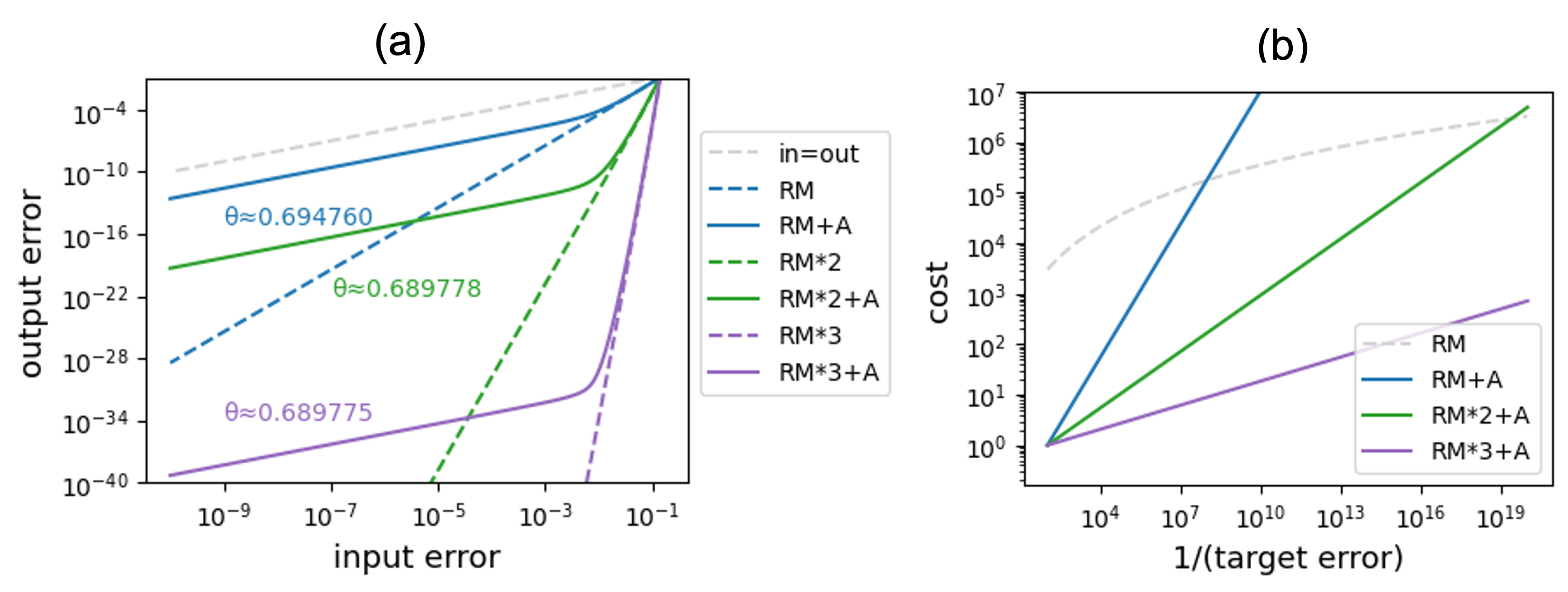}
    \caption{(a) Input-output error relation for different concatenated MSD schemes when executed for one time, with $\epsilon_{in}=0.01$. 'RM' stands for the $[[15, 1, 3]]$ protocols and 'A' stands for the $[[4, 1, 1]]$ protocol. 'RM*t' stands for executing the $[[15, 1, 3]]$ protocol for $t$ times, and 'RM*t+A' stands for executing the $[[15, 1, 3]]$ protocol for $t$ times and then the $[[4, 1, 1]]$ protocol.  The $\theta$ for target magic states is shown near the curve. Although concatenation cannot improve the convergence order, it can significantly increase the prefactor for linear convergence. (b) Estimation of raw state cost for distillation in the asymptotic limit of executions. For the 'RM' protocol, we also considered the overhead for gate synthesis. With more concatenation with the 'RM' protocols, we can reduce the overall cost for distillation.  }
    \label{fig:concat_eff}
\end{figure}

Although all the known exotic MSD protocols only allow linear error suppression and have unsatisfying cost scaling compared with canonical protocols, we may reduce the overhead by concatenating exotic MSD protocols with canonical protocols with good cost scaling. For example, we may concatenate the $[[4, 1, 1]]$ code with the $[[15, 1, 3]]$ code that allows for order-3 error suppression for multiple levels. We may greatly reduce the prefactor for linear convergence as shown in Fig. \ref{fig:concat_eff}(a), though we cannot upgrade the convergence order to more than linear asymptotically. We further estimate the cost for obtaining exotic magic states from our concatenated protcols and the $[[15, 1, 3]]$ protocol together with gate synthesis. The cost for a $n$-to-$k$ MSD protocol with linear error suppression $\epsilon_{out}=k'\epsilon_{in}$ and success probability $p_s$ is given by
\begin{equation}
    C = (\frac{\epsilon_{in}}{\epsilon_{tar}})^{\beta}, \quad \beta = \frac{\log{n/kp_s}}{ \log{1/k'}}
\end{equation}
In comparison, the cost for $[[15, 1, 3]]$ protocol scales with $C_d \propto\log^\gamma(1/\epsilon_{tar})$ where $\gamma = \log_3 15\approx 2.465$. For the optimal gate synthesis based on $T$ gates, the cost $C_g$ just scales with $\log(1/\epsilon_{tar})$ linearly with a constant prefactor \cite{t_gate_count}. We can therefore approximate the total cost with $C_{total}(\epsilon_{tar}) \approx C_g(\epsilon_{tar}) * C_d(\epsilon_{tar}/C_g(\epsilon_{tar})) $ \footnote{The target error for distillation process should be lower than the target error rate as it takes $C_g$ copies of states to synthesis into the target states. As the worst cases estimation, we require the target error for distillation to be $\epsilon_{tar}/C_g(\epsilon_{tar})$ such that the final states after synthesis will meet the target error requirement.}. We plot the scaling of cost in Fig. \ref{fig:concat_eff}(b). Although all of the concatenated exotic MSD protocols cannot outperform the canonical protocol asymptotically, we may still find practical error regime where our protocols may render advantage.
\section{Summary and open questions}

In this work, we proposed a method that maps MSD protocols to dynamical systems using the framework of stabilizer reduction. We showed it to be a convenient way to visualize the process of MSD under iteration, and demonstrate efficiency analysis for common MSD protocols. Besides, we applied our method to those exotic MSD protocols proposed in \cite{howardSmallCodesMagic2016}. We are able to have a further understanding on the condition of distillable exotic magic states and our method can also be used to analyze these procotols in an analytical way. Furthermore, we numerically studied the potential of concatenated MSD schemes using our method. We show that we might discover MSD protocols that distill into various exotic magic states, and the overhead due to linear error suppression can be reduced by further concatenation with efficient MSD protocols.

However, we have to emphasize that MSD protocols still require more understanding, especially for these protocols not based on transversal $T$ gates. Here are some open questions that are worth more effort in our opinion:
\begin{itemize}
    \item Is there any fundamental limitation for a magic state to be distillable? Given an arbitrary angle $\theta$ and arbitrary error tolerance $\epsilon$, can we construct a MSD protocol that distill $\ket{\theta'}$ with $|\theta - \theta'|<\epsilon$ with reasonable code size? If so, gate synthesis might no longer be necessary, and we can always distill into the magic states corresponding to the $R_Z(\theta)$ gates we need.
    \item Can we find exotic MSD protocols with at least quadratic error suppression? By far all exotic MSD protocols we know only suppress error linearly. If we can improve it to at least quadratic, then the overhead can be reduced exponentially.
    \item How can we engineer MSD protocols such that it can distill into a given $\ket{\theta'}$? We have show concatenation has the potential to generate MSD protocols for new exotic magic states. How should we play around it to answer the first question? How many codes we may need, and what's the distribution of all distillable $\theta$ if we restrict our system size to a finite value?
    \item How to study MSD performance using dynamical system mapping without the perfect Clifford assumption? People have been assuming perfect Clifford operations and Pauli measurements in MSD process as they are fault-tolerant and the logical noise can be suppressed with increased code distance. However, for practical MSD with finite-size QEC codes, logical error for Clifford operations still cannot be ignored and they could impact on the final output states. Even though there has been some work targeting on this problem \cite{jochym2013robustness, zheng2025magic},  a more unified and generic understanding on the problem is needed.
\end{itemize}
We hope this work could pave the way for more versatile MSD protocols and deepen the common understanding of the fundamental principle of MSD protocols.

\section*{Code availability}
The code used to produce the figures in this work can be found in this \href{https://github.com/Dran-Z/Mapping-MSD-to-Dynamical-Systems}{Github repository}.

\section*{Acknowledgment}
We thank Yuanchen Zhao and Johannes Borregaard for insightful discussion and Mark Howard for kind encouragement. We thank Pei-Kai Tsai for valuable comments on the manuscript. We thank the anonymous reviewer for various comments that greatly improved the quality of the manuscript. This work was supported by National Natural Science Foundation of China (Grants No. 92365111), Shanghai Municipal Science and Technology (Grant No. 25LZ2600200)  and Beijing Natural Science Foundation (No. Z220002)
\bibliographystyle{quantum}
\bibliography{mapMSD_revise}

\appendix
\section{Algorithm for mapping stabilizer codes}

We here describe the implementation of our mapping algorithm. The input should be a list of stabilizer generators \texttt{generator\_set}, and a list of logical operators \texttt{logical\_operators}. The output will be a list of polynomial function used for describing $p_s(x,y,z)$, $T^{x,y,z}_i(x,y,z)$.
\begin{algorithm}[H]
\caption{Mapping Algorithm for Polynomials}
\label{alg:mapping}
\begin{algorithmic}[1]
\Require \texttt{generator\_set}, \texttt{logical\_operators}
\Ensure $\{p_s(x,y,z), \, T^{x,y,z}_i(x,y,z)\}$
\vspace{0.5em}

\State \textbf{Compute stabilizer set:}
\Statex \hspace{1em}Generate all elements of the stabilizer group by multiplying all possible
combinations of generators in \texttt{generator\_set}. 
Store as \texttt{stabilizer\_set}.

\State \textbf{Construct $p_s(x,y,z)$:}
\Statex \hspace{1em}Initialize dictionary \texttt{counter\_ps}$=\emptyset$.
\For{each stabilizer $s_j \in$ \texttt{stabilizer\_set}}
    \State Count Pauli weights $(w_x, w_y, w_z)$ of $s_j$.
    \If{$(w_x,w_y,w_z) \in$ \texttt{counter\_ps}}
        \State Increment its value by $1$.
    \Else
        \State Add entry $(w_x,w_y,w_z) \mapsto 1$.
    \EndIf
\EndFor
\State Convert \texttt{counter\_ps} into polynomial $p_s(x,y,z)$ by treating each
entry as a monomial with corresponding coefficient.

\State \textbf{Construct $T^{x,y,z}_i(x,y,z)$ for each logical qubit:}
\For{each logical qubit $i$ with logical operators $X_i, Z_i$}
    \For{$\ell \in \{X,Y,Z\}$}
        \State Generate $\ell$\texttt{stabilizer\_set} by multiplying $\ell_i$ with all $s_j \in$ \texttt{stabilizer\_set}.
        \State Initialize dictionary \texttt{counter\_$\ell i$}.
        \For{each element $u \in \ell$\texttt{stabilizer\_set}}
            \State Count Pauli weights $(w_x, w_y, w_z)$ of $u$.
            \State Update \texttt{counter\_$\ell i$} accordingly.
        \EndFor
        \State Convert \texttt{counter\_$\ell i$} into polynomial $T^\ell_i(x,y,z)$.
    \EndFor
\EndFor

\State \textbf{Output:} $\{p_s(x,y,z), T^{x,y,z}_i(x,y,z)\}$ for all logical qubits.
\end{algorithmic}
\end{algorithm}


\section{Description of $[[15, 1, 3]]$ and $[[14, 2, 2]]$ codes}
\label{app: code_description}
Both the $[[15, 1, 3]]$ \cite{bravyiUniversalQuantumComputation2005} and $[[14, 2, 2]]$ \cite{bravyiMagicstateDistillationLow2012} codes have logical transversal $T$ gates and therefore can be used as MSD protocol to distill the $T$ states. We give their stabilizer description here as they are practical small examples to work with.

The $[[15, 1, 3]]$ code has 10 $Z$-type stabilizers and 4 $X$-type stabilizers. It's parity check matrix is given by
\begin{equation}
    H_X = \begin{bmatrix}
    1 & 0 & 1 & 0 & 1 & 0 & 1 & 0 & 1 & 0 & 1 & 0 & 1 & 0 & 1\\ 
    0 & 1 & 1 & 0 & 0 & 1 & 1 & 0 & 0 & 1 & 1 & 0 & 0 & 1 & 1\\  
    0 & 0 & 0 & 1 & 1 & 1 & 1 & 0 & 0 & 0 & 0 & 1 & 1 & 1 & 1\\ 
    0 & 0 & 0 & 0 & 0 & 0 & 0 & 1 & 1 & 1 & 1 & 1 & 1 & 1 & 1
    \end{bmatrix}
\end{equation}
\begin{equation}
        H_Z = \begin{bmatrix}
    1 & 0 & 1 & 0 & 1 & 0 & 1 & 0 & 1 & 0 & 1 & 0 & 1 & 0 & 1\\ 
    0 & 1 & 1 & 0 & 0 & 1 & 1 & 0 & 0 & 1 & 1 & 0 & 0 & 1 & 1\\  
    0 & 0 & 0 & 1 & 1 & 1 & 1 & 0 & 0 & 0 & 0 & 1 & 1 & 1 & 1\\ 
    0 & 0 & 0 & 0 & 0 & 0 & 0 & 1 & 1 & 1 & 1 & 1 & 1 & 1 & 1\\
    0 & 0 & 1 & 0 & 0 & 0 & 1 & 0 & 0 & 0 & 1 & 0 & 0 & 0 & 1\\
    0 & 0 & 0 & 0 & 1 & 0 & 1 & 0 & 0 & 0 & 0 & 0 & 1 & 0 & 1\\
    0 & 0 & 0 & 0 & 0 & 1 & 1 & 0 & 0 & 0 & 0 & 0 & 0 & 1 & 1\\
    0 & 0 & 0 & 0 & 0 & 0 & 0 & 0 & 0 & 1 & 1 & 0 & 0 & 1 & 1\\
    0 & 0 & 0 & 0 & 0 & 0 & 0 & 0 & 0 & 0 & 0 & 1 & 1 & 1 & 1\\
    0 & 0 & 0 & 0 & 0 & 0 & 0 & 0 & 1 & 0 & 1 & 0 & 1 & 0 & 1\\
    \end{bmatrix},
\end{equation}
Notably, $H_Z = [H_X, H_Z']^T$. The logical operator is respectively $X_L = X^{\otimes 15}$, $Z_L = Z^{\otimes 15}$.

The $[[14, 2, 2]]$ code has 9 $Z$-type stabilizers and 3 $X$-type stabilizers. It's parity check matrix is given by
\begin{equation}
    H_X = \begin{bmatrix} 1 & 0 & 1 & 0 & 1 & 0 & 1  &1 & 0 & 1 & 0 & 1 & 0 & 1 \\ 0 & 1& 1&0&0&1&1 & 0 & 1& 1&0&0&1&1 \\ 0&0&0&1&1&1&1 &0&0&0&1&1&1&1\end{bmatrix} \ \ 
    H_Z = \begin{bmatrix} 2&3&4&5\\1&3&4&6\\1&2&4&7\\1&2&8&9\\1&3&8&10\\1&4&8&11\\2&3&11&12\\3&4&8&13\\2&4&8&14 \end{bmatrix},
\end{equation}
where we used the sparse representation for $H_Z$ by writing down the index of non-zero elements, e.g. the first row means $Z_2Z_3Z_4Z_5$. The logical operators are respectively $X_{L,1} = XXXXXXXIIIIIII, Z_{L,1} = ZZZZZZZIIIIIII$, and $X_{L,2} = IIIIIIIXXXXXXX, Z_{L,2} = IIIIIIIZZZZZZZ$.

\section{More flow diagrams }
\label{app: more_flow}
\begin{figure}
    \centering
    \includegraphics[width=\linewidth]{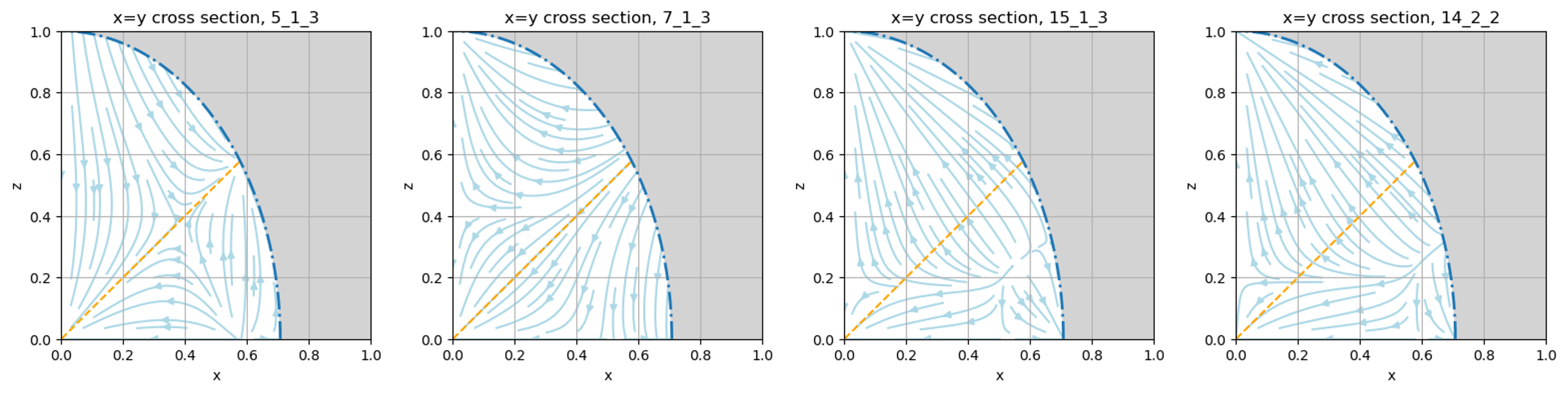}
    \caption{Flow diagram for $x=y$ cross section for the four common protocols. The yellow dashed line denote $x=z$ for reference.}
    \label{fig:xycross}
\end{figure}

In this section, we present more flow diagrams to provide comprehensive understanding for the dynamical systems mapped from the four MSD protocols we discussed in Fig. \ref{fig:ideal_flow}. Firstly, we consider the flow diagram for the $x=y$ cross section. For the $[[5, 1, 3]]$ protocol, the $x=y=z=1/\sqrt{3}$ is shown to be a stable fixed point (Fig. \ref{fig:xycross}(a)) as the protocol can distill into the $\ket{F}$ state.  Both the $[[15, 1, 3]]$ and $[[14, 2, 2]]$ protocols admit the $\ket{T}$ state ($x=y=1/\sqrt{2},z=0$) and the Clifford $\ket{0}$ state as stable fixed points. Notably, the saddle point on the boundary is the state corresponding to $x=y=\cos(\pi/8)/\sqrt{2}, z=\sin(\pi/8)$.

\begin{figure}
    \centering
    \includegraphics[width=\linewidth]{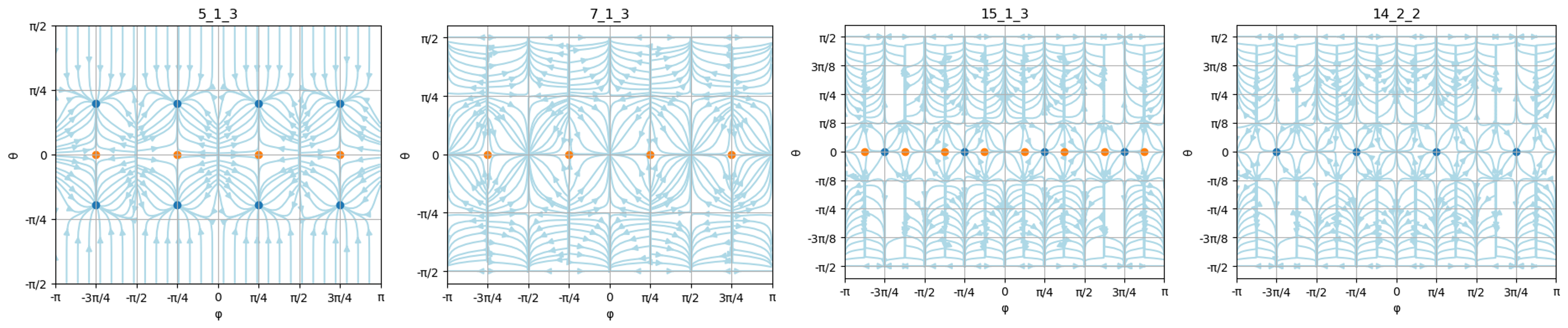}
    \caption{Flow diagram on the Bloch sphere for the four common protocols using Mercator projection. Blue(orange) dots stand for stable(unstable) fixed points for magic states.}
    \label{fig:angular}
\end{figure}

We further consider the distillation dynamics on the Bloch sphere by parametrize the three coordinates with two angular parameters $\theta, \phi$:
\begin{equation}
    x=\cos(\theta)\cos(\phi), y=\cos(\theta)\sin(\phi), z=\sin(\theta).
\end{equation}
In such case, the input states are all pure states. We plotted the flow diagram using the Mercator projection as Fig. \ref{fig:angular}, and all the four protocols exhibit significant symmetry. The $\ket{T}$ state is an unstable fixed point for the $[[5, 1, 3]]$ protocol, and only the $\ket{F}$ state and its Clifford equivalence are stable fixed points. The $\ket{T}$ state for the $[[7, 1, 3]]$ protocol is also unstable on the Bloch sphere. For both the $[[15, 1, 3]]$ and $[[14, 2, 2]]$ protocols, they admit the $\ket{T}$ state and its Clifford equivalence as the stable fixed points and the $\ket{\sqrt{T}}$ state and its Clifford equivalence as the unstable fixed points. 

\begin{figure}
    \centering
    \includegraphics[width=\linewidth]{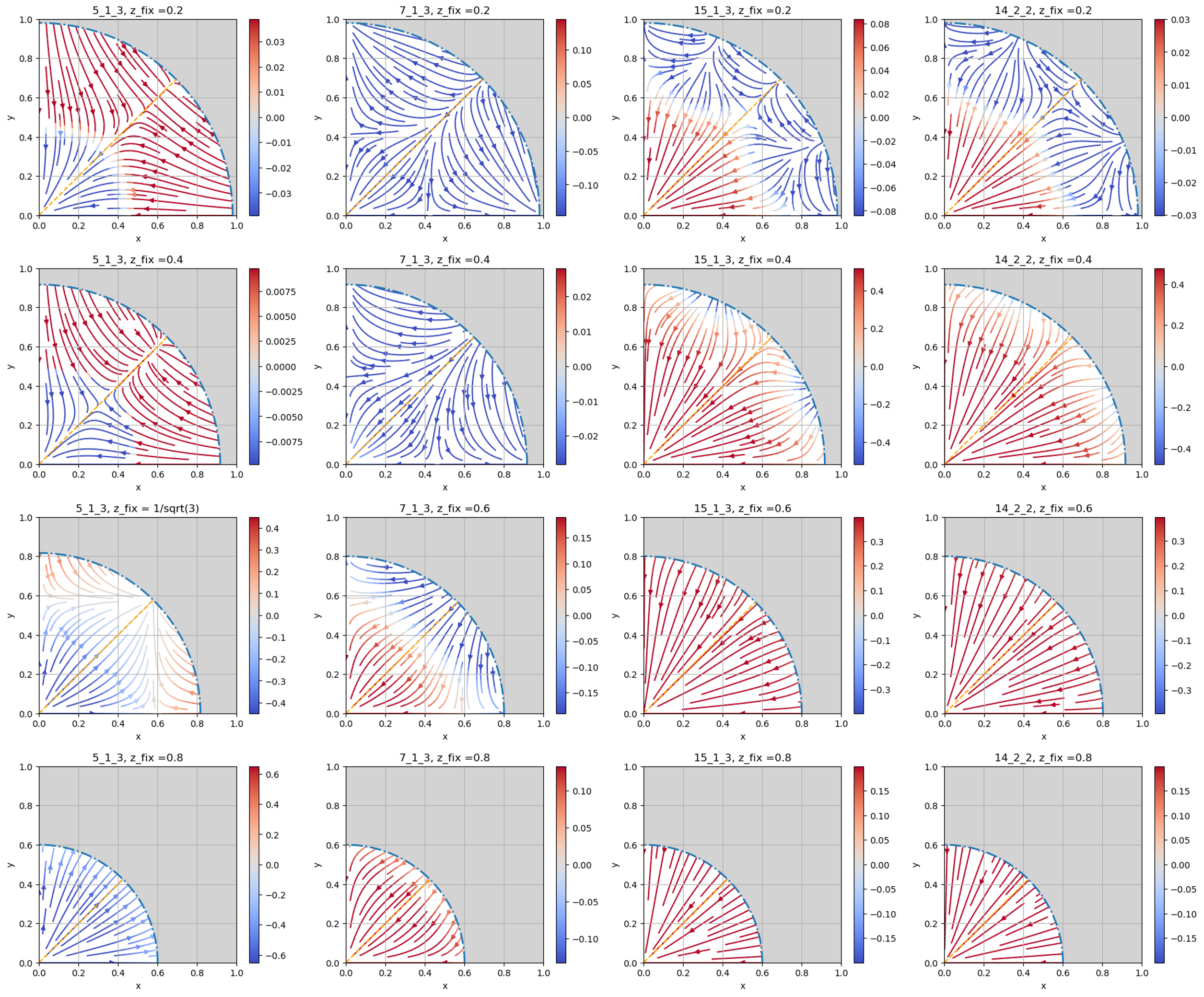}
    \caption{Flow diagram for cross section with $z\neq 0$. Colormap denotes the vertical flow change along the $z$ axis: Blue(red) stands for going down(up).}
    \label{fig:cross}
\end{figure}

We also fixed $z$ for several non-zero values and plotted the cross section flow for the four protocols as shown in Fig. \ref{fig:cross}. As now the cross section is not necessarily a fixed plane, the flow direction will have vertical component and we denote it with heatmap. 


\section{Analytical form of dynamical systems }
\label{app: analytical_descption}
We give the analytical description of the reduced ($z=0$) dynamical system for common codes that produced the Fig. \ref{fig:ideal_flow}. For the $[[15, 1, 3]]$ code, we have
\begin{equation*}
    \begin{cases}
        x'=D_x(x,y) = \dfrac{ x^{15} + 105x^{11}y^4 - 280x^9y^6 + 435x^7y^8 + 15x^7 - 168x^5y^{10} + 35x^3y^{12} + 105x^3y^4}{15x^8 + 15y^8+210x^4y^4 + 1}\\
        y'=D_y(x,y) = \dfrac{ y^{15} + 105y^{11}x^4 - 280y^9x^6 + 435y^7x^8 + 15y^7 - 168y^5x^{10} + 35y^3x^{12} + 105y^3x^4}{15x^8 + 15y^8+210x^4y^4 + 1}
    \end{cases}
\end{equation*}

If we further restrict $x=y$ , we can get
\begin{equation}
    x' = \frac{128x^{15} + 120x^7}{240x^8 + 1}.
\end{equation}
We then set $x=(1-2\epsilon)/\sqrt{2}$ define $\epsilon ' =  (1 -\sqrt{2} x')/2$, which will give
\begin{equation}
    \epsilon' = \frac{1}{2} - \frac{1}{2} \frac{((1-2\epsilon)^{15}+15(1-2\epsilon)^7)}{15(1-2\epsilon)^8+1},
\end{equation}
which matches the analytical result shown in \cite{bravyiUniversalQuantumComputation2005}. 
For the $[[14, 2, 2]]$ codes, the two output states share the same dynamical system:
\begin{equation}
    \begin{cases}
        x'=D_x(x,y) = \dfrac{8x^7+56x^3y^4}{7x^8+98x^4y^4+7y^8+1}\\
        y'=D_y(x,y) = \dfrac{8y^7+56y^3x^4}{7x^8+98x^4y^4+7y^8+1}
    \end{cases}
\end{equation}

For the Steane code,
\begin{equation}
    \begin{cases}
        x' = D_x(x,y) = \dfrac{x^7+7x^3y^4+7x^3}{7x^4+7y^4+1}\\
        y' = D_y(x,y) = \dfrac{y^7+7y^3x^4+7y^3}{7x^4+7y^4+1}.
    \end{cases}
\end{equation}
By setting $x=y$, it is further reduced to $x'=\frac{8x^7+7x^3}{14x^4+1}$, which matches the analytical form provided in \cite{reichardtQuantumUniversalityMagic2005}.
For the $[[5, 1, 3]]$ code,
\begin{equation}
    \begin{cases}
        x' = D_x(x,y) = \dfrac{5xy^2-x^5}{5x^2y^2+1}\\
        y' = D_y(x,y) = \dfrac{5x^2y-y^5}{5x^2y^2+1}
    \end{cases}
\end{equation}

For the four-qubit code,
\begin{equation}
    \begin{cases}
        x' = \dfrac{ x^{2} z^{2} + 2 x z + z^{2}}{ x^{2} z^{2} +  x^{2} + 2 x z + 1}
\\
        z' = \dfrac{ x^{3} + 2 x^{2} z +  x z^{2} +  x}{ x^{2} z^{2} + x^{2} + 2 x z + 1}

    \end{cases}
\end{equation}

For the six-qubit code,
\begin{equation}
    \begin{cases}
        x' = \dfrac{2 x^{4} + 2x^{3} - 2 x^{2} z^{4} + 2 x z^{3} +  z^{4}}{ x^{4} + 2 x^{3} + 4 x^{2} z^{3} + 2 x z^{3} + 1}\\
        z' = \dfrac{ x^{5} z - x^{3} z^{3} + 2 x^{2} z^{2} + 2 x^{2} z +  x z^{3} + 2 x z^{2} +  x z}{ x^{4} + 2 x^{3} + 4 x^{2} z^{3} + 2 x z^{3} + 1}
    \end{cases}
\end{equation}

\section{Fractal property of fixed point distribution}

\begin{figure}
    \centering
    \includegraphics[width=0.9\linewidth]{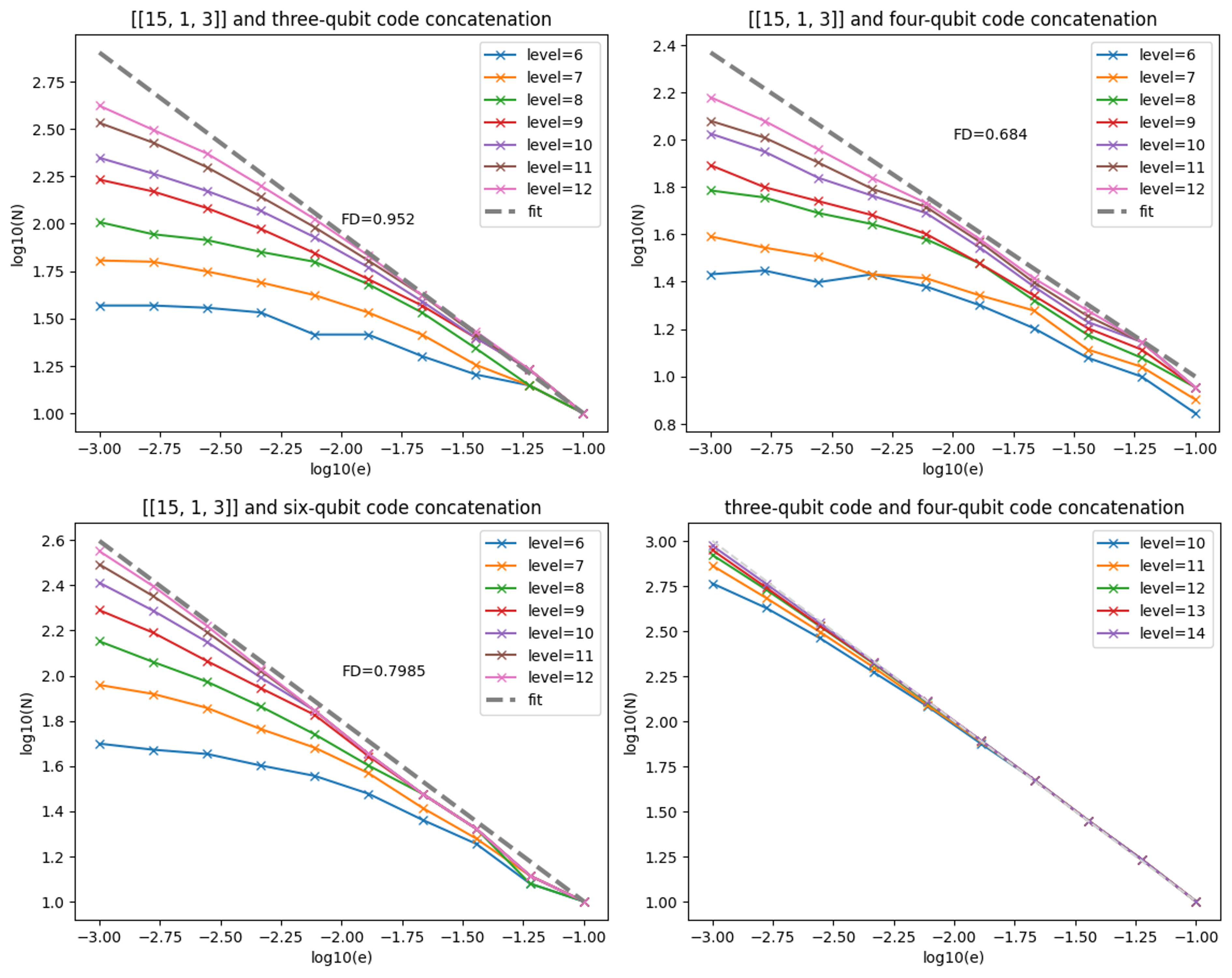}
    \caption{Fractal dimension of the fixed point distribution for concatenated protocols. The fractal dimension is numerically calculated using the box-counting method by varying the size of box length $e$ and counting the number of boxes $N$ needed to cover all points \cite{boxcounting}. Except for the concatenation between the $[[3, 1, 1]]$ and $[[4, 1, 1]]$ code, the other concatenations will give fractal dimension smaller than one.}
    \label{fig:fractal_dimension}
\end{figure}

To verify the distribution of fixed points from concatenated protocols might exhibit fractal property, we exploit the box counting method \cite{boxcounting} to numerically calculate the fractal dimension. We normalize the fixed point in range $[0, 1]$, vary the length of boxes $\epsilon$, and calculate the number of boxes needed to cover all fixed points $N(\epsilon)$. The fractal dimension is given by
$$
F_d = \lim_{\epsilon\rightarrow 0}|\log{N(\epsilon)/\log{\epsilon}}|.
$$
Notice that we are unable to concatenate for arbitrary level to obtain the full distribution of fixed points, we have to truncate $\epsilon$ before the $\log{N}$ saturating. We therefore numerically calculate the fractal dimension by taking a linear fit for the linear part as in Fig. \ref{fig:fractal_dimension}. We find out the fractal dimensions for codes concatenated with $[[15, 1, 3]]$ protocol are all smaller than one, which implies their fractal property in the one-dimensional case. For the concatenation between $[[3, 1, 1]]$ and $[[4, 1, 1]]$, however, the fractal dimension is very close to 1 and the distribution of fixed points might be everywhere but not dense.
\label{app: fractal}

\end{document}